\documentclass[twocolumn, preprintnumbers, 
endnote,nofootinbib,prl,9pt,superscriptaddress]{revtex4}

\usepackage{epsfig,subfigure}

\usepackage{epstopdf}

\usepackage[utf8]{inputenc}
\usepackage{graphicx}
\usepackage{amssymb}
\usepackage{amsxtra}
\usepackage{amsmath}
\usepackage{booktabs,multirow,tabularx}
\usepackage{slashed}
\usepackage{float}
\usepackage{placeins}
\usepackage{rotating}
\usepackage{lscape}
\usepackage{color}
\usepackage{hyperref}
\usepackage{enumitem}

\usepackage[Q=yes,pverb-linebreak=no]{examplep}

\DeclareGraphicsExtensions{.eps}
\graphicspath{{./}}

\newcommand{\pt}{p_{\scriptscriptstyle T}}

\newcommand{\be}{\begin{equation}}
\newcommand{\ee}{\end{equation}}

\newcommand{\mt}{m_t}

\newcommand{\nn}{\nonumber}

\newcommand{\mh}{m_h}

\def\beq{\begin{equation}}
\def\bea{\begin{eqnarray}}
\def\eeq{\end{equation}}
\def\eea{\end{eqnarray}}
\def\beqnl{\begin{align}}
\def\endal{\end{align}}

\makeatletter
\newcommand{\normalorbold}{%
  \ifnum\pdf@strcmp{\math@version}{bold}=\z@ bx\else m\fi
}
\makeatother

\begin{document}

\title{\boldmath 
An Analytical Method for the NLO QCD Corrections to Double-Higgs Production
}

\author{Roberto Bonciani\footnote{email: roberto.bonciani@roma1.infn.it}
}
\affiliation{Sapienza - Universit\`a di Roma, Dipartimento di Fisica, Piazzale Aldo Moro 5, 00185, Rome, Italy}
\affiliation{INFN Sezione di Roma, Piazzale Aldo Moro 2, 00185, Rome, Italy}

\author{Giuseppe Degrassi\footnote{email: degrassi@fis.uniroma3.it}
}
\affiliation{Dipartimento di Matematica e Fisica, Universit{\`a} di Roma Tre,
 I-00146 Rome, Italy}
 \affiliation{INFN, Sezione di Roma Tre, 00146 Rome, Italy}
\author{Pier Paolo Giardino\footnote{email: pgiardino@bnl.gov}
}
\affiliation{Department of Physics, Brookhaven National Laboratory,
Upton, NY 11973, USA}
\author{Ramona Gr\"{o}ber\footnote{email: ramona.groeber@durham.ac.uk}
}
\affiliation{Institute for Particle Physics Phenomenology, Department of Physics, Durham University, Durham, DH1 3LE, UK}

\begin{abstract}
We propose a new method to calculate analytically higher-order perturbative
corrections and we apply it to the calculation of the two-loop virtual corrections to 
Higgs pair production through gluon fusion. The method is based on the expansion of the 
amplitudes in terms of a small Higgs transverse momentum.  This approach gives 
a very good approximation (better than per-mille) of the partonic cross section 
in the center of mass energy region $\sqrt{\hat{s}} \lesssim 750$ GeV, where
$\sim95\%$ of the total hadronic cross section is concentrated.
The presented method is general and can be applied in a straightforward way to
the computation of virtual higher-order corrections to other $2\to2$ processes,
representing an improvement with respect to calculations based on heavy 
mass expansions. 
\end{abstract}

\maketitle

\section{Introduction}

The experimental exploration of the properties of the Higgs boson is
one of the major targets of the Large Hadron Collider (LHC).  However
the self-couplings of the Higgs boson, which in the Standard Model
(SM) are fully determined in terms of the mass of the Higgs boson and
the Fermi constant, have not been probed yet.  While the quartic Higgs
self-coupling is not directly accessible at the LHC
\cite{Plehn:2005nk, Binoth:2006ym}, the trilinear self-coupling might
be measurable from Higgs pair production processes \cite{Djouadi:1999rca, Baur:2003gp,
  Baglio:2012np, Yao:2013ika, Barger:2013jfa, Azatov:2015oxa,
  Lu:2015jza, Dolan:2012rv, Papaefstathiou:2012qe, deLima:2014dta,
  Wardrope:2014kya, Behr:2015oqq, Dolan:2013rja, Dolan:2015zja,
  Englert:2014uqa}.

Those processes, in particular Higgs pair production
in gluon fusion, are also sensitive to new physics, that can greatly
modify their rates \cite{Dawson:2015oha, Dib:2005re, Grober:2010yv,
  Contino:2012xk, Grober:2016wmf}. Bounds on $gg\to HH$ for different
final states are reported in \cite{ Aad:2014yja, Khachatryan:2015yea,
  Aad:2015uka, Aad:2015xja,Khachatryan:2016sey,Aaboud:2016xco}.

Therefore, a precise prediction of the gluon fusion channel is
essential to determine the Higgs trilinear self-coupling and constrain
new physics.  At leading order (LO) the gluon fusion process has been known
since the 80s \cite{Glover:1987nx}.  At next-to-leading order (NLO)
this process is fully known only numerically \cite{Borowka:2016ehy, Borowka:2016ypz},
while analytical results are available in the heavy top mass ($m_t$) limit
 \cite{Grigo:2013rya, Grigo:2015dia, Degrassi:2016vss} and
partially in the light $m_t$ limit \cite{Davies:2018ood}. In
\cite{Grober:2017uho} a method was proposed for obtaining an analytical
result combining large top mass expansion and a threshold expansion by
means of Pad\'e approximants.

The limits of \cite{Grigo:2013rya, Grigo:2015dia, Degrassi:2016vss} and
\cite{Davies:2018ood} well describe the Higgs pair production in the regions
$\sqrt{\hat{s}}<300$ GeV
and $\sqrt{\hat{s}}>750$ GeV respectively,
where $\sqrt{\hat{s}}$ is the partonic center-of-mass energy, but fail
to describe the intermediate region.

We propose a new approach for the analytical calculation of the virtual
NLO corrections to the Higgs pair production through gluon fusion. The
method is based on the expansion of the amplitudes around a small
Higgs transverse momentum $\pt$ and Higgs mass $m_h$. After properly
expanding, the resulting amplitudes are functions of only $m_t$ and
$\sqrt{\hat{s}}$ and can be calculated analytically without resorting to
further expansions. With this method we are able to correctly describe
the Higgs pair production in the region $\sqrt{\hat{s}}\lesssim750$
GeV, nicely complementing the present literature.  It must be also
noted that, due to the shape of the gluon pdfs, this region represents
$~95\%$ of the total hadronic cross section.

Our approach has the virtue of covering larger regions of the phase space
with respect to approaches based on heavy mass expansions 
or high energy expansions and 
can be easily implemented to other $2\to 2$ processes.

In this letter we describe the basics of the method, and the main
results of our calculation, while we will reserve a more detailed
discussion of the computation to future works.

\section{Notation and definitions}

In this section we introduce the notation we will use in the rest of
the paper and define a set of kinematical variables.
The amplitude $g_a^\mu (p_1) g_b^\nu (p_2) \to H(p_3) H(p_4)$ can be written as
\be
A^{\mu \nu} = \frac{G_\mu}{\sqrt{2}} \frac{\alpha_s (\mu_R)}{2 \pi} 
\delta_{ab}\, T_F\, 
\hat{s}\left[ A_1^{\mu \nu}  \,F_1 +  A_2^{\mu \nu}\,  F_2
\right],
\label{eqamp}
\ee
where $G_\mu$ is the Fermi constant, $\alpha_s(\mu_R)$ is the strong
coupling defined at the renormalization scale $\mu_R$ and $T_F=1/2$ is
the normalization factor for the fundamental representation of
$SU(N_c)$. In eq.~\eqref{eqamp} $A_{1,2}^{\mu \nu}$ are the orthogonal
projectors onto the spin-0 and spin-2 states, respectively, while
the corresponding form factors $F_{1,2}$ are functions of $\mt,\: \mh$ and the
partonic Mandelstam variables\footnote{All momenta are assumed incoming.}
\be
\hat{s} = (p_1 + p_2)^2, ~~\hat{t} = (p_1 + p_3)^2,~~
\hat{u} = (p_2 + p_3)^2~,
\label{Mandvar}
\ee
via
\be
F_1 =  F_1 (\hat{s},\hat{u},\mt^2,\mh^2) , \quad
F_2 = F_2 (\hat{s},\hat{u},\mt^2,\mh^2) .
\label{FormFact}
\ee
We defined $A_1^{\mu \nu}$ and $A_2^{\mu \nu}$ as
\bea 
\hspace*{-2mm} A_1^{\mu\nu} &= &
g^{\mu\nu}-\frac{p_1^{\nu}\,p_2^{\mu}}{\left(p_1\cdot p_2\right)} \, ,
\label{A1}\\
\hspace*{-2mm} A_2^{\mu\nu} & = &
- g^{\mu\nu}+ \frac{ m_h^2\, p_1^{\nu}\, p^{\mu}_2}{\pt^2\left(p_1\cdot p_2\right)} 
 \label{A2} \nn\\
&& \hspace*{-10mm} - 2\frac{ 
 \left(p_3\cdot p_2\right)p_1^{\nu}\,p_3^{\mu}
+ \left(p_3\cdot p_1\right)p_3^{\nu}\,p_2^{\mu}
-  \left(p_1\cdot p_2\right) p_3^{\mu}\,p_3^{\nu}}
{\pt^2\left(p_1\cdot p_2\right)} \, , 
\eea 
with $\pt$ the transverse momentum of the Higgs particle, that can
be expressed in terms of the Mandelstam variables as
\be
\pt^2 = \frac{\hat{t}\hat{u} - \mh^4}{\hat{s}}~.
\label{pt2hat}
\ee
The Born cross section, then, is
\be
\sigma^{(0)}(\hat{s})=\frac{G_\mu \alpha_s^2(\mu_R)}{512 (2\pi)^3}
\int^{\hat{t}_+}_{\hat{t}_-}d\hat{t}\left(|T_F F_1|^2+|T_F F_2|^2\right),
\label{Born}
\ee
with $\hat{t}_\pm=-\hat{s}/2(1-2m_h^2/\hat{s}\mp\sqrt{1-4m_h^2/\hat{s}})$.
For our purpose, it is particularly convenient to introduce the prime
Mandelstam variables:
\bea
&&s' = p_1\cdot p_2 = \frac{\hat{s}}{2}, ~~t' = p_1 \cdot p_3 =\frac{\hat{t}-m_h^2}{2},~~\nn\\
&& u' = p_2 \cdot p_3=\frac{\hat{u}-m_h^2}{2}~,
\eea
for which $s'+t'+u'=0$. In these variables the Higgs transverse momentum becomes
\be
\pt^2=2\frac{t' u'}{s'}-m_h^2 \label{pt2}.
\ee

Our ultimate goal is to make an expansion for small $\sqrt{\pt^2 + \mh^2}
\sim \pt$. Since the  final result is symmetrical in $t'\leftrightarrow u'$,
the latter can be achieved expanding\footnote{Expanding only in $t'\sim
  0$ would not be correct if the final result were not symmetrical in
  $t'\leftrightarrow u'$}
 for $t^\prime \sim 0, \: u^\prime \sim -s^\prime$. This is going to restrict 
 $F_1$ and $F_2$ in eq.~(\ref{FormFact}) to a forward kinematic,
 namely to be function of 
$\hat{s}/\mt^2$ only, reducing the computational difficulty from
a three scales problem to a single scale one. To perform
the expansion we need to express the momenta in terms of the parallel and
transverse components w.r.t the beam axis.  For this purpose we define the
combination of momenta
\begin{equation}
r=p_1+p_3 \quad \text{and} \quad \bar{r}=p_2+p_3.
\label{rdef}
\end{equation}
It is easy to show that
\bea
&&r^2=\hat{t},\, \bar{r}^2=\hat{u},\nn\\
&&p_1 \cdot r = -p_2 \cdot r = t'\,, \nn\\
&&p_2 \cdot \bar{r} = - p_1 \cdot \bar{r} = u' \,,
\label{relr}
\eea
and that
\bea
r^{\mu}&=\frac{t'}{s'}( - p_1^{\mu} +  \, p_2^{\mu}) +r^{\mu}_{\perp}\nn\\
\bar{r}^{\mu}&=\frac{u'}{s'}( \, p_1^{\mu} - \, p_2^{\mu} )+\bar{r}^{\mu}_{\perp},
\label{rdef2}
\eea
where $r^\mu_\perp=\bar{r}^\mu_\perp$ is perpendicular to $p_1$ and $p_2$ and, as
expected, 
\be
r_\perp^2= m_h^2+ 2 t' +2\frac{t'^2}{s'}=-\pt^2.
\label{rperp2}
\ee
Finally, in this reparametrization, $A_{1,2}^\mu$ assume particularly simple
forms:
\be 
A_{1}^{\mu\nu} =
g^{\mu\nu}-  \frac{p_1^{\nu}\,p_2^{\mu}}{s'} \, , \quad
A_{2}^{\mu\nu} =
A_1^{\mu\nu}+ 2
\frac{r_{\perp}^{\mu}r_{\perp}^{\nu}}{\pt^2} \, .
\label{A2p}
\ee

\section{Expansion}

From eq.~\eqref{pt2hat}, assuming real valued $\hat{t}$ and $\hat{u}$,
we obtain the condition
\be
\pt^2 +m_h^2 \leq \frac{\hat{s}}{4},
\label{ptls}
\ee 
that allows us to expand for $\pt^2/s'\ll 1$ and $m_h^2/s'\ll1$.

Although our program is clear, it is hindered by the fact that $\pt$
does not appear directly at the amplitude level. However it is possible to
show that an expansion for $r^\mu\sim0^\mu$ is equivalent to an
expansion in $\pt^2\sim0$.  Using
eqs.~(\ref{rdef2}) and (\ref{rperp2}), and noticing that $r_{\perp}$
is purely space-like, we can exchange the expansion in
$\pt^2\sim0$ with an expansion in $r^\mu\sim0^\mu$ or, equivalently,
$p_3^\mu\sim-p_1^\mu$.

This observation is one of the main results of this letter, and it
allows us to proceed.
We can then rewrite the form factors in eq.~(\ref{eqamp}) as
\begin{eqnarray}
F_{1,2}&=& F_{1,2}\bigg|_{p_3=-p_1}  +r^\mu\frac{\partial F_{1,2}}{\partial p_3^{\mu}}
  \bigg|_{p_3=-p_1} \nn\\
&& + \frac12 r^\mu r^\nu\frac{\partial^2 F_{1,2}}
  {\partial p_3^{\mu}\partial p_3^{\nu}}\bigg|_{p_3=-p_1} +\dots 
\label{exp}
\end{eqnarray}

Although eq.(\ref{ptls}) is always valid, the expansion proposed in (\ref{exp})
requires an hierarchy between $r^2$ and $\mt^2$. 
We are going to estimate the range of validity of the small $\pt$ expansion by
comparing at the LO the result obtained via eq.~\ref{exp} with the exact LO 
result (see next section).

We conclude this section with an important remark on how to correctly
truncate the series in eq.~(\ref{exp}). Since the final result should
be symmetrical for $p_1\leftrightarrow p_2$ and $\partial
F_{1,2}/\partial p_3^{\mu}$ is a rank 1 tensor, the second term in
eq.~(\ref{exp}) r.h.s can be rewritten as $r^\mu \mathcal{F}_{1,2}
(p_1^\mu +p_2^\mu)=\mathcal{F}_{1,2} (\hat{t} + \hat{u})$, with
$\mathcal{F}_{1,2}$ a function of $s'$ and $m_t^2$. For similar
arguments, the third term should be instead proportional to $r^\mu
r^\nu (g_{\mu\nu}+\dots)=\pt^2+\dots$. It is clear, then, that to
expand to the first order in $\pt^2$ one has to expand to the second
order in $p_3^\mu$, or, more in general, an order $n$ expansion in
$\pt^2$ needs the order $2n$ expansion in $p_3^\mu$.

\section{Computation and results}

We generated the relevant amplitudes for the virtual NLO corrections
to $gg\to HH$ with FeynArts \cite{Hahn:2000kx}. The amplitudes were
contracted with the two orthogonal projectors eq.~\eqref{A1} and
\eqref{A2}, using FeynCalc \cite{Shtabovenko:2016sxi}, and reduced to
a combination of scalar integrals. The integrals were then Taylor
expanded, as described in the previous section. Subsequently, the resulting
integrals were reduced in terms of
a basis of Master Integrals (MI) using FIRE \cite{Smirnov:2014hma} and
LiteRed \cite{Lee:2013mka}. All of the MIs, of which nearly the 
totality can be expressed in terms of multiple polylogarithms, were 
already known in the literature
\cite{Bonciani:2003te,Bonciani:2003hc,Aglietti:2006tp,Anastasiou:2006hc,Becchetti:2017abb,Caron-Huot:2014lda,vonManteuffel:2017hms}. However, we evaluated them again directly 
in the phase space region of interest. We cross-checked our results using 
SecDec \cite{Borowka:2015mxa}.  
The details of the calculation 
presented
here, as well as a detailed study of the validity of our
  approximation at the hadronic level, will be the topic of a second paper on this argument
\cite{prep}, while in this letter we will focus on the final result.

In order to show that our method correctly describes the partonic cross 
section for $\sqrt{\hat{s}}<750$ GeV, we will start applying it to the LO.

\begin{figure}
\hspace*{-0.5cm}\includegraphics[width=0.55\textwidth]{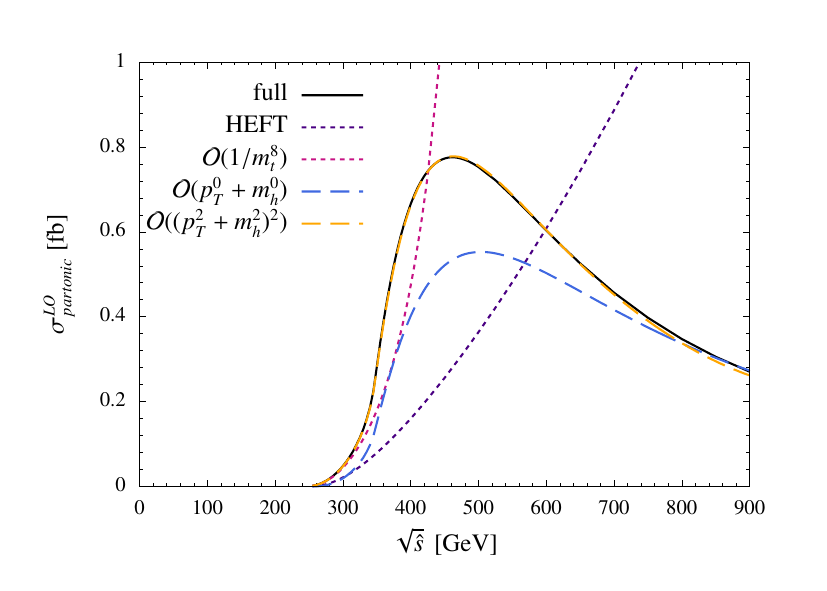}
 \caption{Partonic cross section of $gg\to HH$ as a function of the
   partonic center of mass energy. The black continuous line is the
   full result \cite{Glover:1987nx}. The dotted lines represent two
   orders of approximation in the heavy $m_t$ limit. The dashed lines
   are the result of the small $\pt^2$ approximation presented in this
   letter.}
 \label{XSLO}
\end{figure}

In fig.~\ref{XSLO} we report our calculation for the partonic cross
section using eq.~\eqref{exp}. As discussed in the introduction,
while the heavy $m_t$ expansion describes well only the range
$\sqrt{\hat{s}}<2 m_t$, with our method we are able to correctly
describe a wider range.
It is also interesting to note that an expansion up to
order $\mathcal{O}(\pt^4)$
is already sufficient to describe the complete result with enough precision.

The range of validity of the small $\pt$ expansion can be estimated 
comparing the partonic cross section calculated with our method with 
the one from the full LO calculation.

\begin{table}[ht]
  \centering
  \begin{tabular}{|c||c|c|c|c|c|c|}
  \hline
$\Delta\sigma$ | $\hat{s}$&  $4 m_t^2$   & $6 m_t^2$  & $8 m_t^2 $ &  $12 m_t^2$   &  $16 m_t^2 $ &  $32 m_t^2$\\ \hline \hline
$\pt^0\times 10^{-1}  $    & 6.2 & 4.4 & 3.2 & 1.8 & 1.0 & 0.3 \\ \hline
$\pt^2\times 10^{-2}$      & 8.5 & 4.4 & 1.1 & 2.4 & 5.1 & 33.2 \\ \hline
$\pt^4\times 10^{-2}$   & 1.3 & 0.1 & 0.4 & 0.2 & 0.9 & 2.8\\ \hline
$\pt^6\times 10^{-3}$ & 2.3 & 0.9 & 1.0 & 0.1 & 3.5 & 450 \\ \hline
  \end{tabular}
  \caption{Relative difference between the approximated and the exact LO cross sections, for different orders of expansion, at various $\hat{s}$.}
\label{tab:1}
\end{table}

In table~\ref{tab:1}, we show 
\be
\Delta\sigma=\left|\frac{2(\sigma_{\text{full}}-\sigma_{\text{exp}})}{(\sigma_{\text{full}}+\sigma_{\text{exp}})}\right| ,
\ee
where $\sigma_{\text{full}}$ is the cross section calculated without expansion,
and $\sigma_{\text{exp}}$ is the one calculated in this letter. The table
indicates that $\Delta \sigma$ is small and very well under control up to
values of the partonic c.m. energy of about $\sim 750$ GeV. 
Moreover, in the region of interest, the approximation rapidly improves as one 
considers higher order in the expansion in $\pt^2$ and $m_h^2$. 
The range of validity of our formulas is complementary to the one present in 
the literature, and represents $~95\%$ of the total hadronic cross section.

\begin{figure}
\hspace*{-0.5cm}\includegraphics[width=0.55\textwidth]{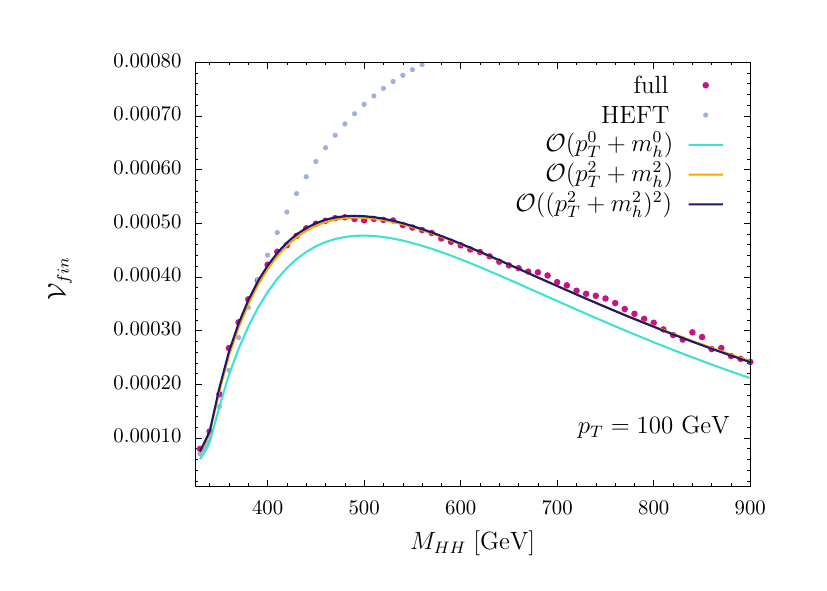}
 \caption{Finite part of the virtual corrections as a function of the invariant mass of the two Higgs system. The pink points are
   extracted with the interpolation function from
   \cite{Heinrich:2017kxx}. The dotted light blue points correspond to
   reweighted HEFT \cite{Dawson:1998py}. The solid lines are the
   respective orders in our calculation. We do not show
   $\mathcal{O}((p_T^2+m_h^2)^3)$ as the line lies perfectly on top of
   the one of $\mathcal{O}((p_T^2+m_h^2)^2)$. }
 \label{NLO}
\end{figure}

This behaviour is confirmed (and even improved) in the comparison with the full numerical result at NLO.
It is well known that the NLO virtual corrections are IR divergent and
these divergences cancel against the ones that come from real
corrections \cite{Grigo:2013rya, Grigo:2015dia,
  Degrassi:2016vss}. Following \cite{Degrassi:2016vss}, we cancel the
IR divergences by adding the counterterm $1/(2\epsilon^2)
F_{1,2}^{LO}(\epsilon^2)(\hat{s})^{-\epsilon}$, where
$F_{1,2}^{LO}(\epsilon)$ are the LO form factors with the inclusion of
the $\mathcal{O}(\epsilon,\epsilon^2)$ terms.  In fig.~\ref{NLO} we
compare our result to the numerical results from
\cite{Borowka:2016ehy}, at the partonic level, using the grid and the interpolation function
for the finite part of the virtual corrections $\mathcal{V}_{fin}$
provided in \cite{Heinrich:2017kxx}. As can be inferred from the
figure, our expansion perfectly agrees with the full result when
the first correction  in $p_T$ and $m_h$ is included. It can
clearly be seen that our lines smooth out the error on the full result
stemming from the interpolation. Furthermore, we compare the numerical results
of the authors of Ref.\cite{Borowka:2016ehy} with ours in several  points of
the grid provided by  the same authors.
For $p_T \lesssim 200$ GeV we find agreement
between the two computations  within  the error
quoted for each point of the grid from the numerical integration. For larger
$p_T$ the agreement is still
quite good (for $p_T \lesssim 300$ GeV is within twice the numerical error
quoted) 
showing a  degradation  with the increase of $p_T$.

\section{Conclusion}
In this letter we have proposed a novel approach for the analytical
computation of the NLO virtual corrections to Higgs pair production
through gluon fusion. This method, based on a expansion for small
$\pt^2$, allows us to describe accurately the region $\hat{s}\lesssim750$
GeV that until now has been explored only numerically.  In particular
we showed that a few terms in the expansion already reproduce the full
LO within $10^{-3}$, in the region of interest. At NLO we find
excellent agreement already at $\mathcal{O}(p_T^2+m_h^2)$ comparing to
the full result of \cite{Borowka:2016ehy}.
To judge the usefulness of our analytic method we compare the CPU time
needed to produce a phase-space point in our approach with that needed in the
numerical calculation of Ref\cite{Borowka:2016ehy}. In order  to compute one
single phase-space point Ref.\cite{Borowka:2016ehy} quotes
an average of 2 hours per node using 16 Dual NVDIA
TESLA K20X GPU nodes  while in our approach the computation of
one single phase-space point took $\sim 4$  seconds on a MacBook Air.
We remark  that this method is general and can be useful for the analytic
computation of radiative corrections to other fundamental
processes for the physics programme of the LHC. In particular the application of this method to the computation of the
  NLO
  virtual corrections to the top contribution in the $HZ$, $ZZ$ and
  $\gamma \gamma$
gluon fusion production processes is expected to be straightforward, while
processes where the top and bottom contribution cannot be separate, like e.g.
in the $WW$ gluon fusion production, deserve a more detailed investigation.

\section*{Acknowledgements}
P.P.G. would like thank Sally Dawson for insightful discussion.
The work of P.P.G. is supported by the U.S. Department of
Energy under Grant Contracts DE-SC0012704. R.G. would like to thank
Stefan Jahn and Johannes Schlenk for help with SecDec and the 
INFN, Sezione di Roma Tre and the theory group of LNF for their hospitality.
R.G. is supported by a European Union COFUND/Durham 
Junior Research Fellowship under the EU grant number 609412.

\bibliographystyle{utphys}
\bibliography{BDGG}

\end{document}